\documentclass[twocolumn,notitlepage,superscriptaddress]{revtex4-1} 

\usepackage{graphicx}
\usepackage{colortbl}
\usepackage{amsmath,amssymb,bm}

\begin{document}

\title{Surface plasmon-mediated nanoscale localization of laser-driven sub-THz spin dynamics in magnetic dielectrics}

\author{Alexander L. Chekhov}
\email{chekhov@shg.ru}
\affiliation{Department of Physics, Moscow State University, 119991 Moscow, Russia}
\affiliation{Fritz Haber Institute of the Max Planck Society, 14195 Berlin, Germany}
\author{Alexander I. Stognij}
\affiliation{Scientific-Practical Materials Research Centre of the NASB, 220072 Minsk, Belarus}
\author{Takuya Satoh}
\affiliation{Department of Physics, Kyushu University, 819-0395 Fukuoka, Japan}
\author{Tatiana V. Murzina}
\affiliation{Department of Physics, Moscow State University, 119991 Moscow, Russia}
\author{Ilya Razdolski}
\email{razdolski@fhi-berlin.mpg.de}
\affiliation{Fritz Haber Institute of the Max Planck Society, 14195 Berlin, Germany}
\author{Andrzej Stupakiewicz}
\affiliation{Faculty of Physics, University of Bialystok, 15-245 Bialystok, Poland}

\date{\today}

\maketitle \textbf{
Ultrafast all-optical control of spins with femtosecond laser pulses is one of the hot topics at the crossroads of photonics and magnetism with a direct impact on future magnetic recording. Unveiling light-assisted recording mechanisms for an increase of the bit density beyond the diffraction limit without excessive heating of the recording medium is an open challenge. 
Here we show that surface plasmon-polaritons in hybrid metal-dielectric structures can provide spatial confinement of the inverse Faraday effect, mediating the excitation of localized coherent spin precession with 0.41 THz frequency. 
We demonstrate a two orders of magnitude enhancement of the excitation efficiency at the surface plasmon resonance within the 100 nm layer in dielectric garnet. Our findings broaden the horizons of ultrafast spin-plasmonics and open pathways towards non-thermal opto-magnetic recording at the nano-scale.
}

The development of next generation magnetic memory devices will be driven by the demand for fast magnetization switching, low energy consumption and high-density recording \cite{KirilyukRMP10}. Although addressing the foremost issue, all-optical magnetization switching with femtosecond laser pulses in various metallic systems \cite{StanciuPRL07, RaduNature11, OstlerNatCom12, ManginNatMat15} requires heating close to the Curie temperature. Recently, new routes of the non-thermal magnetization reversal in a dielectric garnet with the fastest ever magnetic recording event accompanied by extremely low heat load were demonstrated \cite{StupakiewiczNat17}. It is now widely accepted that the future of high-density all-optical magnetic recording depends on the achievements of sub-diffractional nanophotonics \cite{ChallenerNatPhot09,StipeNatPhot10,ChooNatPhot12,LiuNanoLett15, SchmisingNJP15} featuring light localization on the nanoscale by surface plasmon resonances. As such, fundamental understanding of the interactions of high-frequency coherent spin dynamics with plasmonic excitations on both nanometer and sub-picosecond scales in opto-magnetic media is highly desirable.

\begin{figure*}[t]
\centerline{\includegraphics[width=0.8\textwidth]{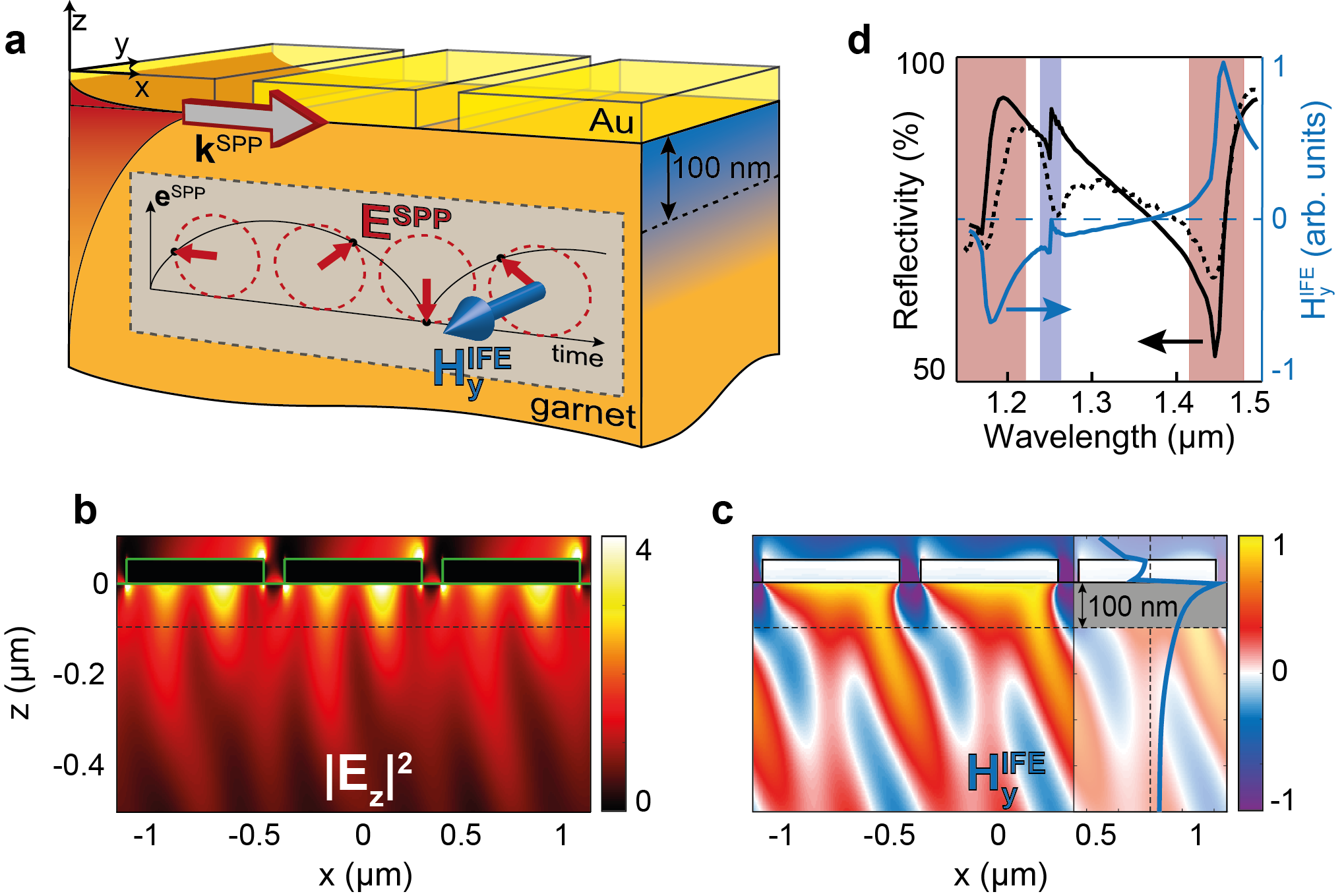}}
\caption{{\bf Surface plasmon-driven IFE confinement.} a) Experimental concept of the SPP-induced IFE. Effective static magnetic field induced by a propagating SPP at the Au/magnetic garnet interface. Neglecting the losses, the latter has two electric fields components, $E_x$ parallel to the SPP wavevector ${\bf k}^{\rm SPP}$, and $E_z$ perpendicular to the interface. The $\pi/2$ phase shift between the two projections of the electric field gives rise to $H^{\rm IFE}_y$ homogeneously distributed in-plane. The spatial localization of the SPP electric field results in the interfacial confinement of $H^{\rm IFE}_y$ on the scale of $100$ nm (governed by the optical properties of the dielectric medium). Numerical simulations of the spatial distributions of the SPP electric field $|E_z|^2$ (b) and effective magnetic field $H^{\rm IFE}_y$ (c) at the SPP resonance at 1450 nm wavelength and 34 degrees angle of incidence. The inset in (c) illustrates the spatial localization of the laterally-averaged $H^{\rm IFE}_y$ in the garnet layer ($z<0$) on the scale of $100$ nm. (d) Spectral dependence of the reflectivity $r$ (solid black line) and SPP-induced effective magnetic field $H^{\rm IFE}_y$ (blue) calculated using the Finite Difference Time Domain method. Experimental reflectivity spectrum is shown with a black dashed line.}
\label{model}
\end{figure*}

\section*{Theory}

Enabling excitation of spin eigenmodes by circularly polarized femtosecond light pulses via the inverse Faraday effect (IFE)
\cite{Kimel05,HansteenPRB06,Kalashnikova07,Reid10,ParchenkoAPL13}, the interaction of the spin excitations and light (optical field ${\bf E}$) is described by the free energy $\mathcal{F}\propto-\delta\varepsilon_{ij} E_i E^*_j$, where $\delta\varepsilon_{ij}$ is the variation of the dielectric permittivity $\varepsilon$ by magnons. Introducing a phase shift $\varphi$ between $E_i$ and $E_j$, for the resulting projection of the effective magnetic field $H^{\rm IFE}_k$ we get: 

\begin{equation}\label{efffield}
    H^{\rm IFE}_k=-\frac{\partial\mathcal{F}}{\partial M_k}\propto \varepsilon_{ij}(E_iE^*_j-E^*_iE_j)\propto \varepsilon_{ij}\lvert E \rvert^2 \sin\varphi.
\end{equation}
This field exerts a torque on magnetization ${\bf M}$ and triggers its precessional motion. Since for circularly polarized light in transparent media $\varphi=\pm\pi/2$, the sign of $H^{\rm IFE}$ is governed by the helicity of light polarization. For any linear polarization $\varphi=0$ and the IFE vanishes. Microscopically, bearing a notable similarity \cite{Battiato14} to the second-order optical rectification, an impulsive stimulated Raman scattering (ISRS) \cite{Yan85} employs two optical fields at frequencies $\omega_1$, $\omega_2$ contained in the spectrum of the femtosecond laser pulse for the excitation of a spin eigenmode at $\Omega=\omega_1-\omega_2$ \cite{Kalashnikova08,Gridnev08}.

The explicit phase dependence in Eq.~(\ref{efffield}) reveals its key role for the ultrafast optomagnetic phenomena. Modern nanophotonics provides effective tools for engineering electric fields in nano-confined volumes, enabling resonant control of their amplitudes and phases. The spin-photon coupling can thus be mediated by collective electronic excitations, i.e. plasmons. The excitation of propagating surface plasmon polaritons (SPPs) at the metal-dielectric interface imposes a rigid phase shift $\varphi$ between the in-plane $x$ (parallel to the SPP propagation direction) and out-of-plane $z$ components of the electric field \cite{Raether}: $E_z/E_x= ik^{\prime}_x/k^{\prime\prime}_z$. Here $k^{\prime}_x$ is the real part of the SPP wavenumber $ k^{\rm SPP}=\frac{\omega}{c}\sqrt\frac{\varepsilon_m\varepsilon_d}{\varepsilon_m+\varepsilon_d}=k^\prime_x+ik^{\prime\prime}_x$, $k^{\prime\prime}_z=\frac{\omega}{c}\frac{\varepsilon_d}{\sqrt{\varepsilon_m+\varepsilon_d}}$, where $\varepsilon_m$, $\varepsilon_d$ refer to the permittivities of the metal and the dielectric, respectively. It is seen that although the SPPs are excited with linearly polarized light, in the SPP TM-wave there is a $\varphi=\pi/2$ phase shift between $E_x$ and $E_z$. Notably, despite the oscillatory spatial dependence of these fields $\propto e^{ik_x^\prime x}$, the phase shift remains constant, giving rise to the static effective magnetic field $H^{\rm IFE}_y$ (Fig.~1a):

\begin{equation}
    H^{\rm IFE}_y\propto\varepsilon_{ij}\lvert E_x\rvert^2 \frac{k^{\prime}_x}{k^{\prime\prime}_z}e^{-2(k^{\prime\prime}_xx-k^{\prime\prime}_zz)}.
\end{equation}
Owing to the nanoscale SPP field localization and amplification, $H^{\rm IFE}_y$ is enhanced and strongly confined in the vicinity of the metal-dielectric interface, which is otherwise unattainable in transparent dielectric media. The inherently nonlinear-optical origin of $H^{\rm IFE}_y$ enables its twice as strong spatial localization, as compared to the SPP electric fields (Fig.~\ref{model},b-c). The sign of $H^{\rm IFE}$, governed by the helicity of the light polarization in IFE, is now determined by the SPP propagation direction. Recently, a similar mechanism was suggested to drive nonlinear self-modulation of SPPs at magnetic interfaces \cite{Herrmann}.

These considerations are supported by the results of numerical simulations performed for the SPP excited at a model metal/dielectric interface (see Methods). It is seen that the SPP excitation prominently enhances electric fields $E_x$, $E_z$ in the dielectric medium (Fig.~\ref{model}b) and enables sizeable effective magnetic field $H^{\rm IFE}_y$ localized in the 100 nm-thick dielectric layer adjacent to the interface (Fig.~\ref{model}c). Exerting a torque on $M_x$, a short (given by the SPP lifetime, i.e. sub-ps) SPP-driven burst of $H_y$ can be observed experimentally as a SPP-mediated excitation of the coherent spin precession in the dielectric employing {\it p}-polarized light, i.e. when the direct IFE is inactive.

\section*{Experiments}

\begin{figure*}[ht]
\centerline{
\includegraphics[width=0.95\textwidth]{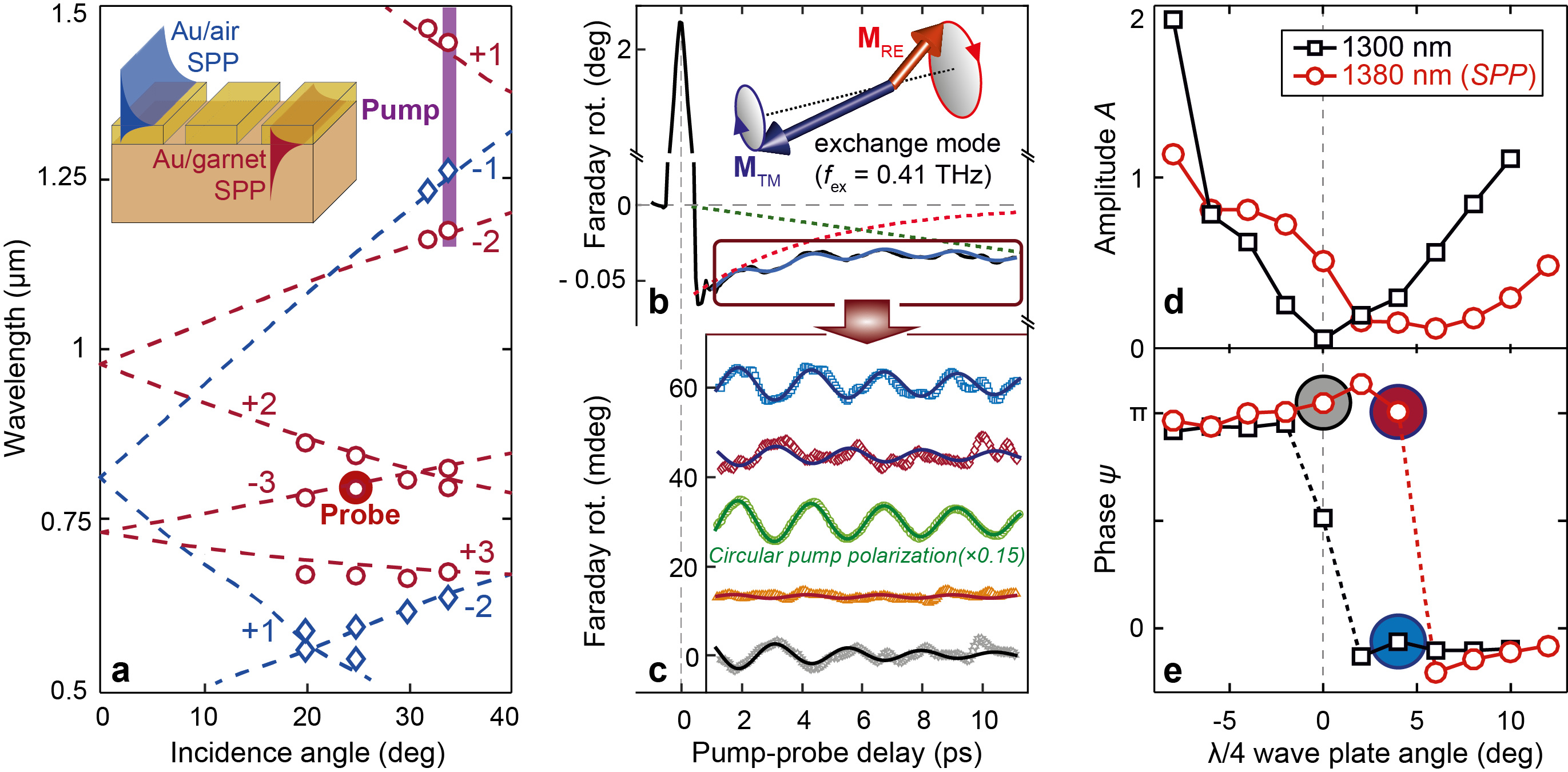}}
\caption{{\bf SPP-mediated excitation of exchange resonance mode.} a) SPP dispersion map of a Au/garnet hybrid magneto-plasmonic crystal with a spatial periodicity of 800 nm, grooves depth of 50 nm and width of 100 nm. The dashed lines indicate the calculated SPP dispersion with the numbers corresponding to the diffraction order. The purple vertical stripe and the red circle show the spectral tunability and width of the pump and probe pulses, respectively, as well as their angles of incidence in the experiments. b) Typical transient variations of the Faraday rotation. Red and green dashed lines indicate background contributions obtained within the data fitting procedure (see Methods). The blue line illustrates the resulted fit to the experimental data. The inset exemplifies the exchange resonance dynamics of the magnetic moments of the rare earth (RE) and transition metal (TM) subsystems. c) The experimental data with the background removed together with the fit lines. The green data points illustrate the spin dynamics excitation with circularly polarized light for comparison purpose. d,e) Amplitude $A$ and phase $\psi$ as a function of the incident pump polarization, as extracted from the fitting procedure. The $\lambda/4$ wave plate angle of $0$ degrees corresponds to the $p$-polarized excitation. The coloured circles correspond to the data shown in (c) with respective colours. A clear shift of $\approx 4^{\circ}$ between the resonant (1380 nm) and off-resonant (1300 nm) highlights the importance of the SPP-driven contribution to the IFE effective magnetic field.}
\label{dispersion}
\end{figure*}

We have complemented a 380 $\mu m$-thick Gd-Yb-doped bismuth iron garnet (GdYbBIG) crystal with a periodically perforated 50 nm-thick Au overlayer, allowing for the excitation of the SPPs at both Au/air and Au/garnet interfaces \cite{Belotelov11,Krutyanskiy15,RazdolskiACS15} (Fig.~\ref{dispersion}a). In a pump-probe scheme, intense $50$ fs near-IR pump pulses excite the SPP inducing an effective static magnetic field ${\bf H}^{\rm IFE}$, which drives a high-frequency ($f_{\rm ex}=0.41$ THz) exchange resonance mode in GdYbBIG \cite{ParchenkoAPL13}. The microscopic excitation mechanism is similar to that discussed in Ref.~\cite{Reid10}. The spin dynamics is monitored via transient Faraday rotation of a delayed, 800 nm probe beam coupled to a SPP at the Au/garnet interface (Fig.~\ref{dispersion}a). In order to verify the SPP role in the excitation of the exchange resonance mode, we performed comparative measurements at the SPP resonance and away from it, taking advantage of the tunability of the pump wavelength (Fig.~\ref{dispersion}a). Figure \ref{dispersion}b shows the characteristic transient variations of the Faraday rotation. Having removed the background, we fitted a decaying sine function to the oscillatory part of the data (Fig.~\ref{dispersion}c) and extracted the fit parameters. Figure~\ref{dispersion}d-e shows the amplitude $A$ and phase $\psi$ of the oscillations as a function of the pump polarization set by a $\lambda/4$ wave plate. For the off-resonant case (black), a parabolic shape of $A$ and a step in $\psi$ is registered, consistent with the usual IFE mechanism. However, at the SPP resonance (red) the $A$ shape distortion and the $\psi$ step shift of about $4^{\circ}$ indicates a strong SPP-induced contribution coexisting with the one originating from the helicity of the pump polarization.

\section*{Data analysis}

\begin{figure*}[t]
\centerline{\includegraphics[width=0.75\textwidth]{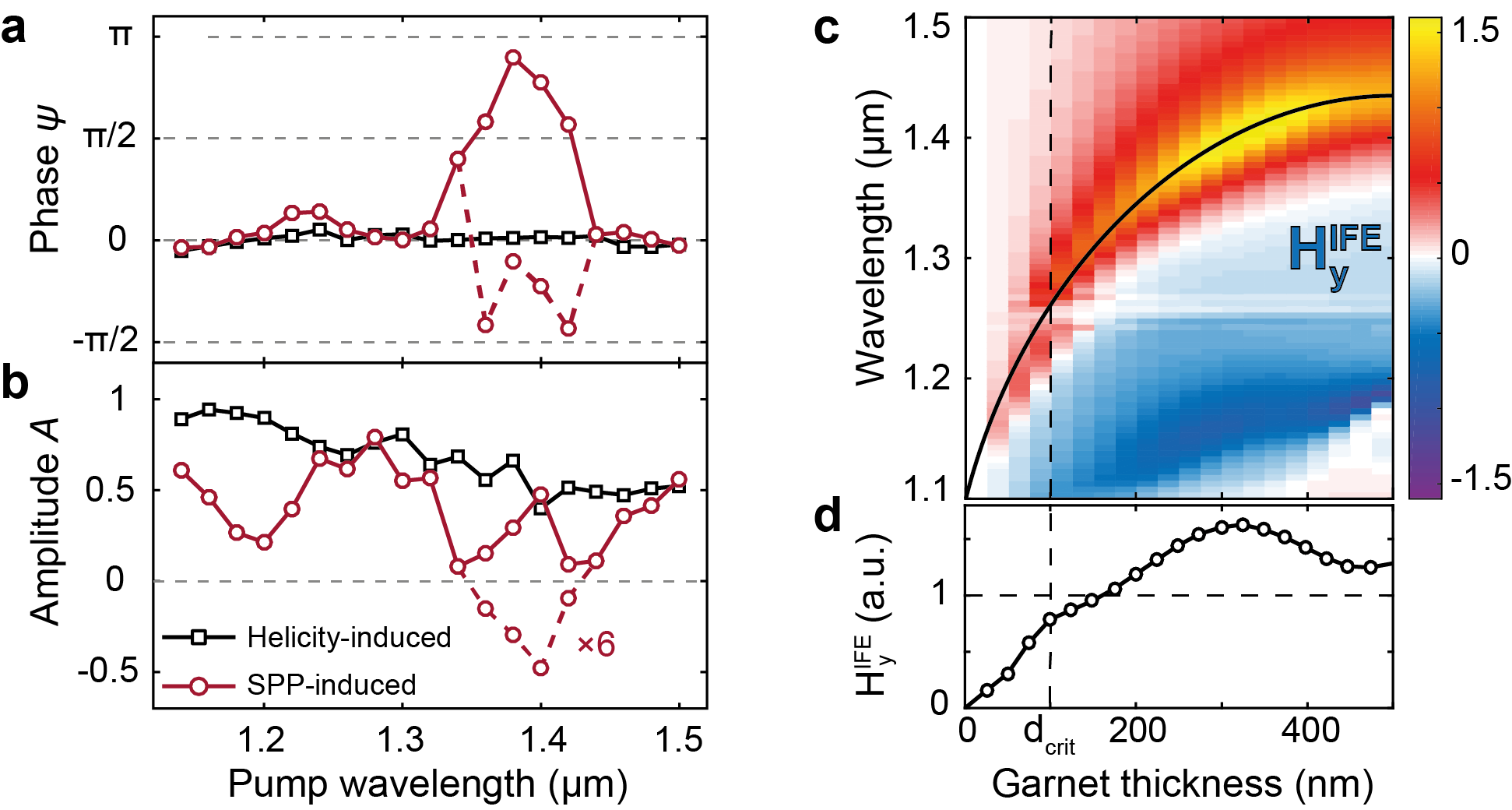}}
\caption{{\bf Spectroscopy of the SPP-mediated IFE.} a-b) Spectral dependencies of the amplitude $A$ and phase $\psi$ of the exchange resonance mode for the circular (black) and close-to-linear (red, 4$^\circ$ $\lambda$/4 plate) pump excitation. The absence of sharp features for the circularly polarized pump pulse hints at negligible variations of the magneto-optical coefficients of the garnet in the spectral region of interest. A prominent feature at about $1400$ nm corresponds to the SPP excitation at the Au/garnet interface. A smaller feature at about $1200$ nm is pertinent to the interplay of the SPP at the Au/air interface and a $2^{\rm nd}$ order SPP at the Au/garnet interface. A phase shift of $\pi$ is mathematically identical to the amplitude sign reversal, which is illustrated by red dashed curves. c-d) Calculated efficiency of the SPP-driven excitation of the exchange resonance mode in thin garnet films. The $H^{\rm IFE}_y$ values are normalized to that obtained for the semi-infinite garnet film. The solid black line in (c) shows the SPP dispersion shift. The dashed vertical line at a garnet thickness of $d_{\rm crit} = 100$ nm indicates the threshold below which the SPP-mediated IFE in hybrid metal-dielectric magneto-plasmonic systems becomes inefficient.}
\label{spectra}
\end{figure*}

In order to unambiguously associate this additional IFE contribution to SPP excitation we performed spectral measurements at a close-to-linear pump polarization where both helicity- and SPP-induced contributions are of comparable magnitude. The spectral dependencies of $A$ and $\psi$ of the oscillations (Fig.~\ref{spectra}a-b) for the circular pump polarization (black) show no significant variations across a broad range of wavelengths. In striking contrast, the $\psi$ spectrum under illumination with a close-to-linearly polarized pump exhibits a large resonant peak centered at around $\lambda_{\rm pump}=1380$ nm, corresponding to the SPP excitation at the Au/garnet interface. The SPP-mediated excitation mechanism is further corroborated by a similar resonant modulation in the $A$ spectrum (Fig.~\ref{spectra}b). Numerical simulations reveal excellent correlation between the SPP excitation and non-zero $H^{\rm IFE}_y$ as well as a good agreement between the simulated and experimental reflectivity spectra (Fig.~\ref{model}d).

With this data in hand, the importance of the SPP-induced contribution to the IFE excitation of the coherent exchange resonance mode is confirmed. Notably, opposite to the homogeneous action of a circularly polarized pump beam in transparent GdYbBIG, the SPP-mediated excitation is localized in the $d_{\rm SPP}\approx 100$ nm-thick layer adjacent to the  interface (see Fig.~\ref{model}c). Considering the Faraday rotation $\theta_F$ proportional to the effective thickness of the optically active medium $d$ (in the homogeneous case $d_{\rm bulk}=380$ $\mu$m), and estimating the SPP-driven precession amplitude to be $\approx10\%$ of the one excited with the circular polarized pump (cf.~Fig.~\ref{spectra}b), for the SPP enhancement of the excitation efficiency $\sigma$ one gets:

\begin{equation}
    \sigma=\frac{A_{\rm SPP}}{A_{\rm bulk}}\frac{d_{\rm bulk}}{d_{\rm SPP}}\approx 4\cdot 10^2.
\end{equation}
Originating in the prominent SPP-driven increase of the electric fields in a dielectric medium, this two orders of magnitude enhancement can be further improved by photonic optimization of the hybrid metal-dielectric system.

\section*{Discussion}

In light of recently discovered magnetic recording via coherent spin precession \cite{StupakiewiczNat17}, strong SPP-mediated confinement of the excitation of coherent spin dynamics holds high promise for increasing the recording density in thin dielectric films. In order to look at the in-depth confinement of ${\bf H^{\rm IFE}}$ upon reducing the GdYbBIG thickness, we simulated spectra of the SPP-induced ${\bf H^{\rm IFE}}$ in thin GdYbBIG films. Illustrating the effective SPP-induced magnetic field $H^{\rm IFE}_y$, Fig.~\ref{spectra},c-d indicate the critical garnet thickness $d_{\rm crit}$ of about 100 nm: below $d_{\rm crit}$, $H^{\rm IFE}_y$ and the quality of a potential SPP-based transducer deteriorates quickly, mostly due to the smearing out of the SPP dispersion (Supplementary Note 1). A model magneto-plasmonic system analyzed here serves as a starting point towards tailoring the local interaction of spins with photons for industrial needs. For instance, stronger confinement of the SPP-mediated excitation can be achieved in transparent media with even higher dielectric function $\varepsilon_d$, paving a way for further material optimization. Approaching the commercially attractive dimensions of a single bit, the demonstrated confinement mechanism holds great potential for future switching and recording applications, as well as sheds light on the fundamental spatio-temporal aspects of the spin-plasmon coupling on ultrashort timescales. These results break new ground in laser-induced coherent spin dynamics, enabling spatial localization of the excitation and thus opening the path towards high-density, low-loss opto-magnetism.

\section*{Methods}

\noindent {\bf Sample fabrication and characterization.} A 380 $\mu$m-thick Gd$_{4/3}$Yb$_{2/3}$BiFe$_5$O$_{12}$ monocrystal with (111) surface orientation was grown by THE liquid phase epitaxy method. A 100 nm-thick gold layer with lateral dimensions of $50\times200 \ \mu$m was deposited on the garnet using the double ion-beam sputter-deposition technique. Then, a periodic lattice of grooves (depth 50 nm, width 100 nm, period 800 nm) was formed in the gold layer by focused ion beam etching. After that, the sample was etched with a defocused ion beam in order to reduce the gold layer thickness down to 50 nm. The fabrication procedure is described in detail in Ref.~ \cite{Chekhov2015}. The magnetic anisotropy field in the GdYbBIG crystal was $0.62$ kOe, the Curie temperature was $T_C= 573$~K and the Gilbert damping was $\alpha= 0.02$, as obtained from the ferromagnetic resonance (FMR) spectroscopy \cite{Satoh2012}.
\\

\noindent {\bf Reflectivity spectroscopy.} The linear reflectivity spectra were obtained using $p$-polarized light focused onto the sample using a lens with $F=5$~cm. The reflected light intensity was registered with a spectrometer (Avesta). The reflectivity data obtained for the sample were normalized to that measured for a non-perforated thick gold film. In the visible range (from $0.5$ to $1$ $\mu$m), a halogen lamp was used as a light source, while for the near-IR wavelengths (from $1$ to $1.5$ $\mu$m) we used a Nd:YAG seeded optical parametric oscillator (Solar LS).
\\

\noindent {\bf Numerical simulations.} The numerical simulations were performed using commercial FDTD software (Lumerical 8.0, www.lumerical.com) on a structure with the same dimensions of the gold grating as in the studied sample (800 nm period, 100 nm groove width, 50 nm thickness) and with optical constants taken from Ref.~\cite{Johnson1972}. Refractive index of the garnet (GdYbBIG) was slightly varied in the range of possible values \cite{Doormann1984} to match the experimentally measured reflectivity spectrum. In the studied spectral region (from 1100 to 1500 nm) it was set to a constant value of $n=2.37$. The light source was polarized in the plane of incidence ($xz$) and the incidence angle was set to $34^{\circ}$ to match the experimental conditions. The GdYbBIG layer was semi-infinite in the $z$-direction perpendicular to the sample surface. In the two other directions the periodic boundary conditions were chosen. The calculated complex values of the electric field components $E_x(x,z)$ and $E_z(x,z)$ at the mesh points were used to obtain the distributions of the effective magnetic field $H_y^{\rm IFE}(x,z)\propto\operatorname{Im}(E_xE_z^*-E_x^*E_z)$ for various wavelengths. The proportionality coefficient is omitted here as we are only interested in the relative strength of $H_y^{\rm IFE}$.

Supplementary Fig. 1a shows the numerically calculated wavelength dependence of the reflectivity and the value of the effective field averaged over the $x$-coordinate (along the interface) at $z=-5$ nm. The effective field spectrum has a maximum at $\lambda=1450$ nm and a minimum at $\lambda=1180$ nm, which corresponds to excitation of Au/garnet SPPs in the $+1$ and $-2$ diffraction orders, respectively. Different propagation directions of these modes result in the opposite signs of the SPP-induced effective magnetic field. Also, the higher diffraction order of the second mode is responsible for the lower quality of the resonance and thus the lower absolute value of the effective field. Supplementary Fig. 1b,c show the effective field distributions for these two resonant wavelengths. Simulations of the SPP-induced IFE in thin GdYbBIG films are summarized in Supplementary Note 1.
\\

\noindent {\bf Time-resolved measurements.} The pump-probe measurements were performed using a Ti:Sa ultrafast amplifier (Ace, SpectraPhysics) with 800 nm output. The pump beam wavelength was converted in an optical parametric amplifier (TOPAS, SpectraPhysics) into wavelengths from 1100 to 1500 nm. A delay line (Newport) in the pump channel was used to control the delay time between the pump and probe pulses. Angles of incidence for the probe and the pump beams were $25^{\circ}$ and $34^{\circ}$, respectively. A polarizer (glan-laser) and a $\lambda$/4 wave plate were used to continuously control the polarization state of the pump beam from linear to circular. The intensity ratio between the probe and the pump beams was about $1:100$. The periodicity of the gold grating was oriented perpendicularly to the plane of incidence, and an external in-plane magnetic field of 1.2 kOe was applied along the SPP propagation direction with the help of an electromagnet. The pump fluence was set to 10 mJ/cm$^2$. The polarization state of the transmitted probe beam was registered in a balanced detection scheme using a lock-in amplifier. A sketch of the experimental scheme is shown in Supplementary Figure 2.
\\

\noindent {\bf Fitting the experimental time-resolved data.} In order to extract the phase $\psi$ and amplitude $A$ of the exchange resonance mode from the experimental data, we performed the fitting taking into account various contributions to the transient Faraday rotation. Characteristic experimental dependence of the transient variations of the Faraday rotation $\Delta\theta_F$ on the pump-probe delay $t$ is exemplified in Supplementary Figure 3a. The data were fitted for the delay times $t>1$ ps after the pump pulse arrival with the following function:
$$
\begin{aligned}
\Delta\theta_F(t)= & a_1\exp(-t/\tau_1)+a_2t+\\
                   & A\exp(-t/\tau_2)\sin(2\pi f_{\rm ex} t+\psi),
\end{aligned}
$$
\noindent where $a_1, \tau_1, a_2, A, \tau_2, \psi$ are the fitting parameters, and the frequency $f_{\rm ex}=0.41$ THz is shared for all the datasets. The first (exponential) term with an amplitude $a_1$ and a decay time $\tau_1$ is pertinent to the heat dissipation at the Au/garnet interface. The second term with the amplitude $a_2$ represents a contribution from the FMR precession mode (precession of the net GdYbBIG magnetization, Supplementary Figure 3b) \cite{ParchenkoAPL13}. For the $1.2$ kOe external magnetic field, its frequency was found to be $f_{\rm FMR}\approx 3$ GHz. Since the HF precession frequency is more than two orders of magnitude larger than that of the FMR precession, at the timescales of interest in this Manuscript the FMR mode-driven variations of the Faraday rotation can be approximated with a linear function. Finally, the last term in the fit function describes the decaying exchange resonance mode with a frequency $f_{\rm ex}$ and decay time $\tau_2$ (Supplementary Figure~3c).

\section*{Acknowledgements}
The authors thank M. Wolf and A. Maziewski for continuous support. We acknowledge support from the National Science Centre Poland (grant DEC-2017/25/B/ST3/01305) and G-RISC grant 2017a-18.

\onecolumngrid
\renewcommand{\figurename}{SUPPLEMENTARY FIGURE}
\setcounter{figure}{0}
\clearpage

\section*{Supplementary Note 1. Numerical simulations for thin garnet films}

In order to investigate the possibility of the SPP-mediated excitation of spin dynamics in thin GdYbBIG films, we performed calculations of the field distribution for the same parameters as described in Methods, but with finite garnet thicknesses $d$. The garnet layer was placed between the Au layer and the semi-infinite gallium gadolinium garnet (GGG) layer with a refractive index of $n=1.95$. Supplementary Fig. 4a shows the calculated reflectivity map versus the wavelength and the GdYbBIG layer thickness. The reflectivity minimum indicated with the black dashed curve corresponds to the excitation of the Au/garnet SPP. At small $d$, the SPP electric field feels the presence of the lower-index GGG layer, and the reflectivity minimum experiences a blue shift. In the limit of $d=0$, the Au/garnet SPP becomes an Au/GGG SPP, and the effective magnetic field vanishes. 

Further, the feature associated with the excitation of the Au/air SPP is indicated with the white dashed line. As expected, the dispersion of this SPP mode does not shift with the variations of $d$. Additional sharp features that shift with the garnet thickness correspond to the excitation of the transverse-magnetic waveguide (TM-WG) modes in the garnet layer~[1]. These bulk modes naturally disappear at small $d$.

The probed amplitude of the spin precession excited by virtue of the SPP is proportional to the average $H_y^{\rm IFE}$ inside the garnet layer. To that end, we have integrated the calculated effective magnetic field $H_y^{\rm IFE}(x,z)$ over the $x$-axis and over the $z$-axis in the range $-0.5 \ \mu$m $<z<0 \ \mu$m for $d>0.5 \ \mu$m and $-d<z<0 \ \mu$m for $d\leqslant0.5 \ \mu$m. The resulting integrated effective magnetic field dependence normalized to its maximal value in the case of semi-infinite garnet layer is plotted in Supplementary Fig. 4b. The strongly pronounced maximum corresponds to the effective magnetic field induced by the Au/garnet SPP. We note that although the TM-WG modes demonstrate sizeable effective magnetic field too, they disappear at small $d$ and are not localized at the Au/garnet interface, unlike the SPPs.

The largest values of the integrated effective magnetic field at the Au/garnet SPP resonance are plotted (as a cross-section) in Supplementary Fig. 4c. The periodical modulation at $d>0.3 \ \mu$m is associated with the self-interference of the pump beam upon multiple reflections in the thin GdYbBIG film. As the garnet layer thickness becomes comparable with the SPP penetration depth in the dielectric medium ($\approx 100$ nm), the integrated effective magnetic field starts to decrease quickly with $d$. Additional variations at small $d$ occur due to the crossing of the two SPP dispersion curves.
\\

[1] Chekhov, A.L. et al. Wide tunability of magnetoplasmonic crystals due to excitation of multiple waveguide and plasmon modes. {\it Opt. Express} {\bf 22} 17762-17768 (2013)

\newpage
\begin{figure*}[htb]
\centerline{\includegraphics[width=0.65\textwidth]{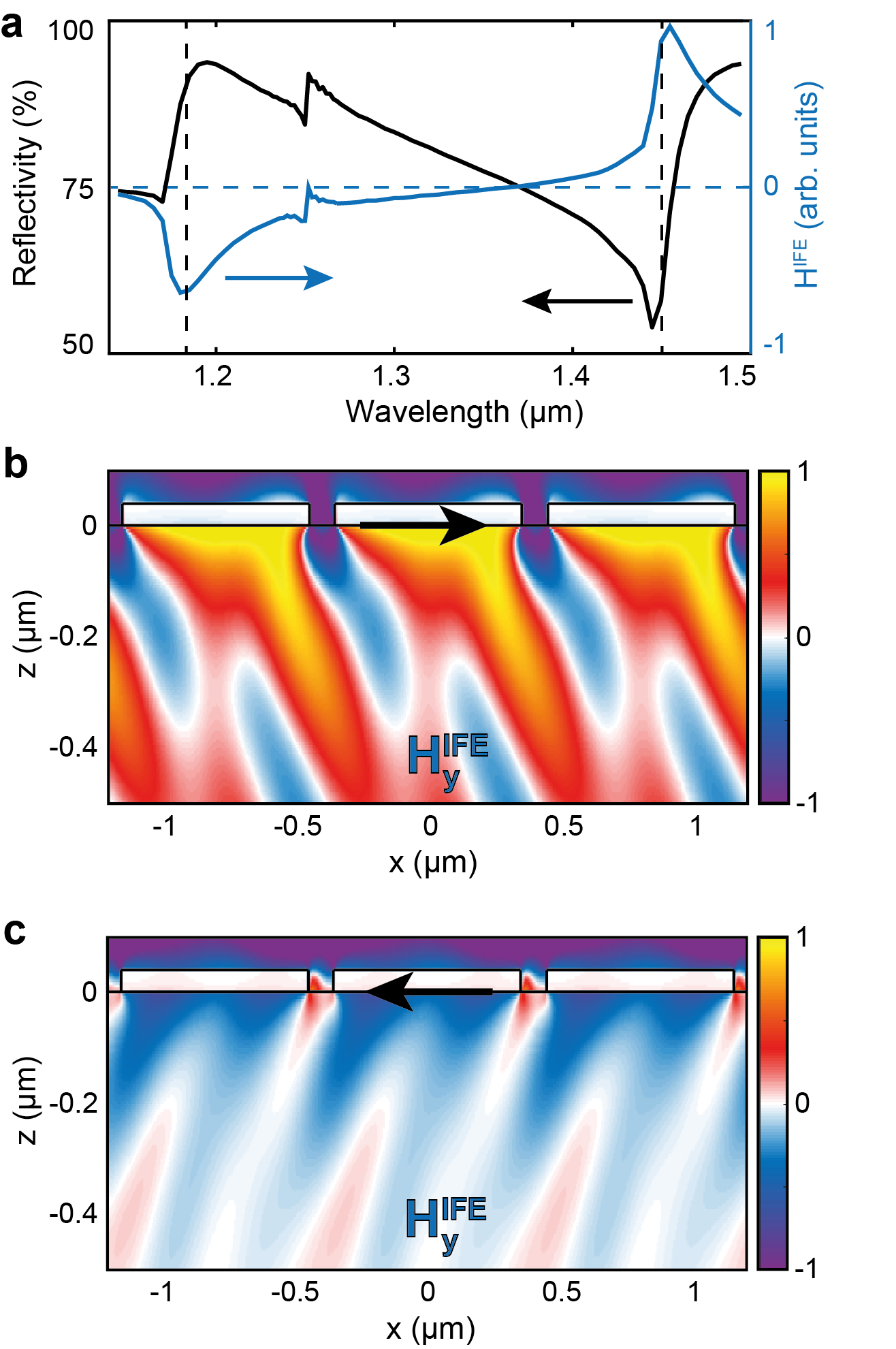}}
\caption{(a) Calculated spectra of the reflectivity (black) and effective magnetic field (blue). (b,c) Effective magnetic field distributions calculated at wavelengths of 1280 and 1450 nm, respectively. The arrows at the Au/garnet interface indicate the direction of the SPP propagation in each case.}
\label{SupFig1}
\end{figure*}

\newpage
\begin{figure*}[htb]
\centerline{\includegraphics[width=0.65\textwidth]{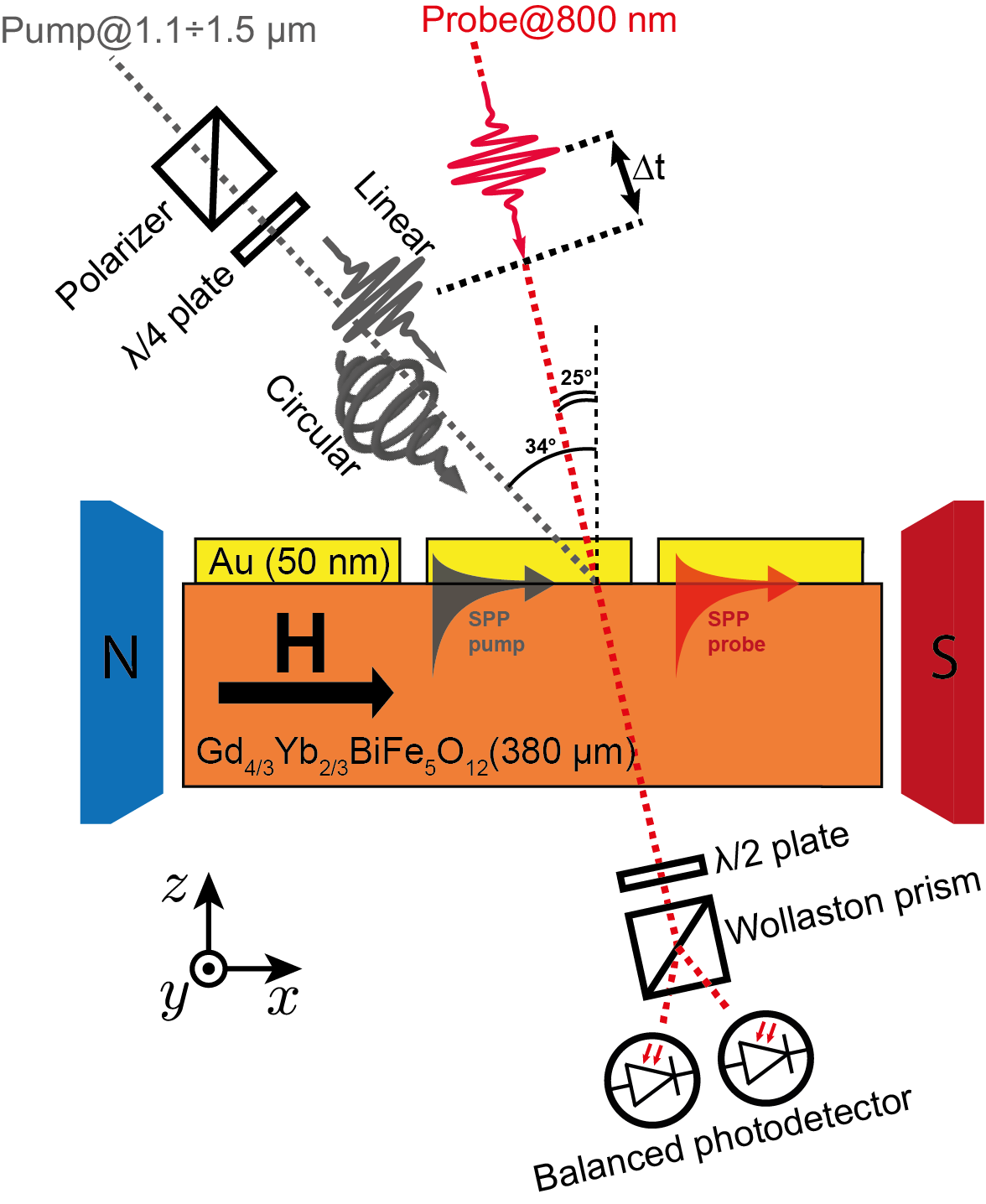}}
\caption{Schematics of the experimental setup. Pump and probe beams were incident under 34 and 25 degrees, correspondingly. Linearly polarized probe excited a SPP on Au/garnet interface. Pump polarization state was controlled with a polarizer (Glan-laser) and a $\lambda$/4 wave plate. Linearly polarized pump excited a SPP on the same interface as well. External magnetic field of 1.2 kOe was applied parallel to the SPP propagation direction. Transient variations of the probe beam Faraday rotation were measured using a balanced photodetection scheme with a lock-in amplifier. Time delay $\Delta$t between the pump and probe pulses was changed using a delay line in the probe channel.}
\label{SupFig2}
\end{figure*}

\newpage
\begin{figure*}[htb]
\centerline{\includegraphics[width=0.6\textwidth]{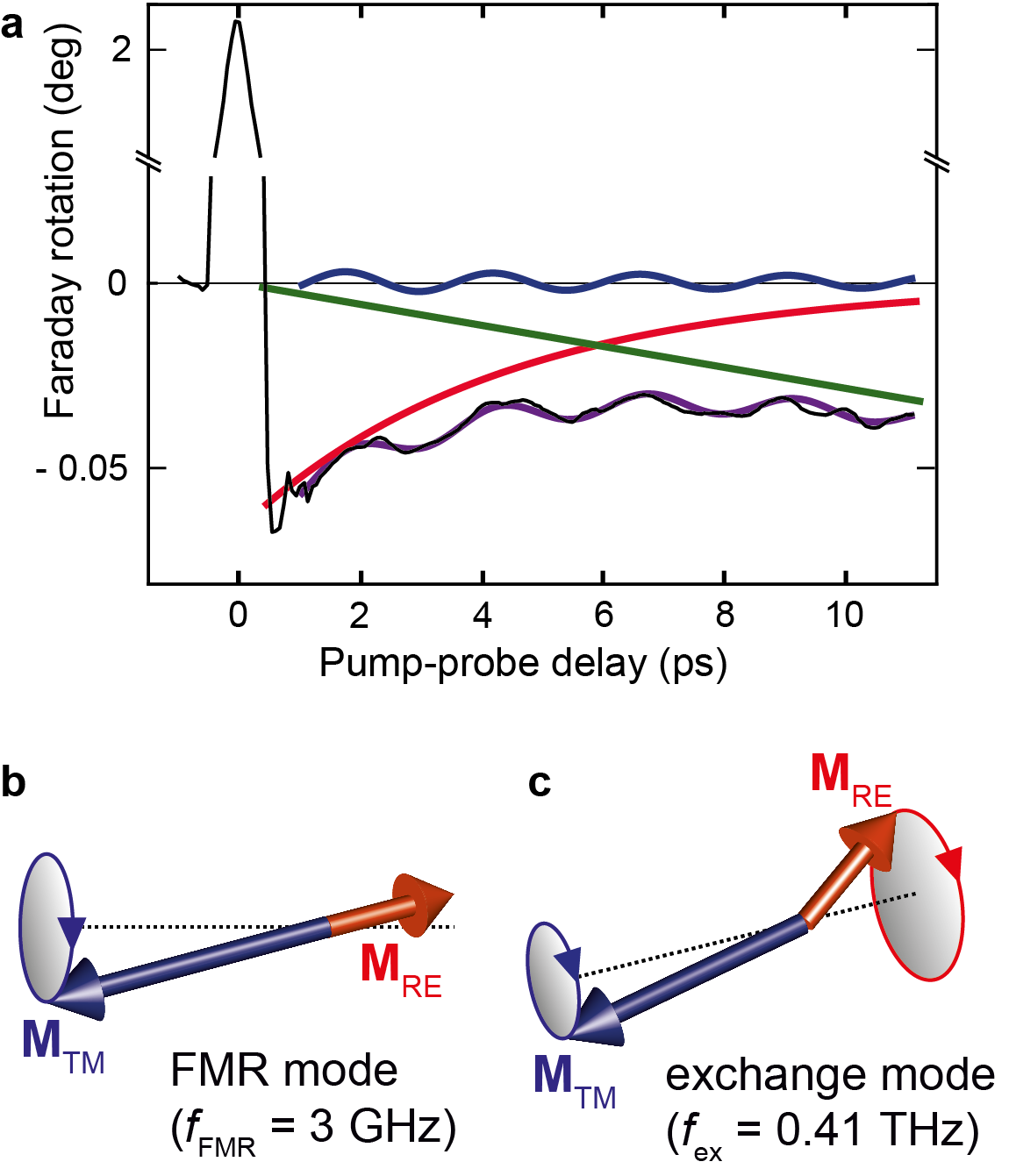}}
\caption{(a) Experimentally measured Faraday rotation versus the pump-probe delay (black curve) and different contributions used in the fitting function: heat dissipation (red solid line), FMR precession (green solid line), exchange resonance mode (blue solid line), and the sum of all contributions (purple solid line). (b,c) Illustrations of the FMR and exchange resonance modes.}
\label{SupFig2}
\end{figure*}

\newpage
\begin{figure*}[htb]
\centerline{\includegraphics[width=0.65\textwidth]{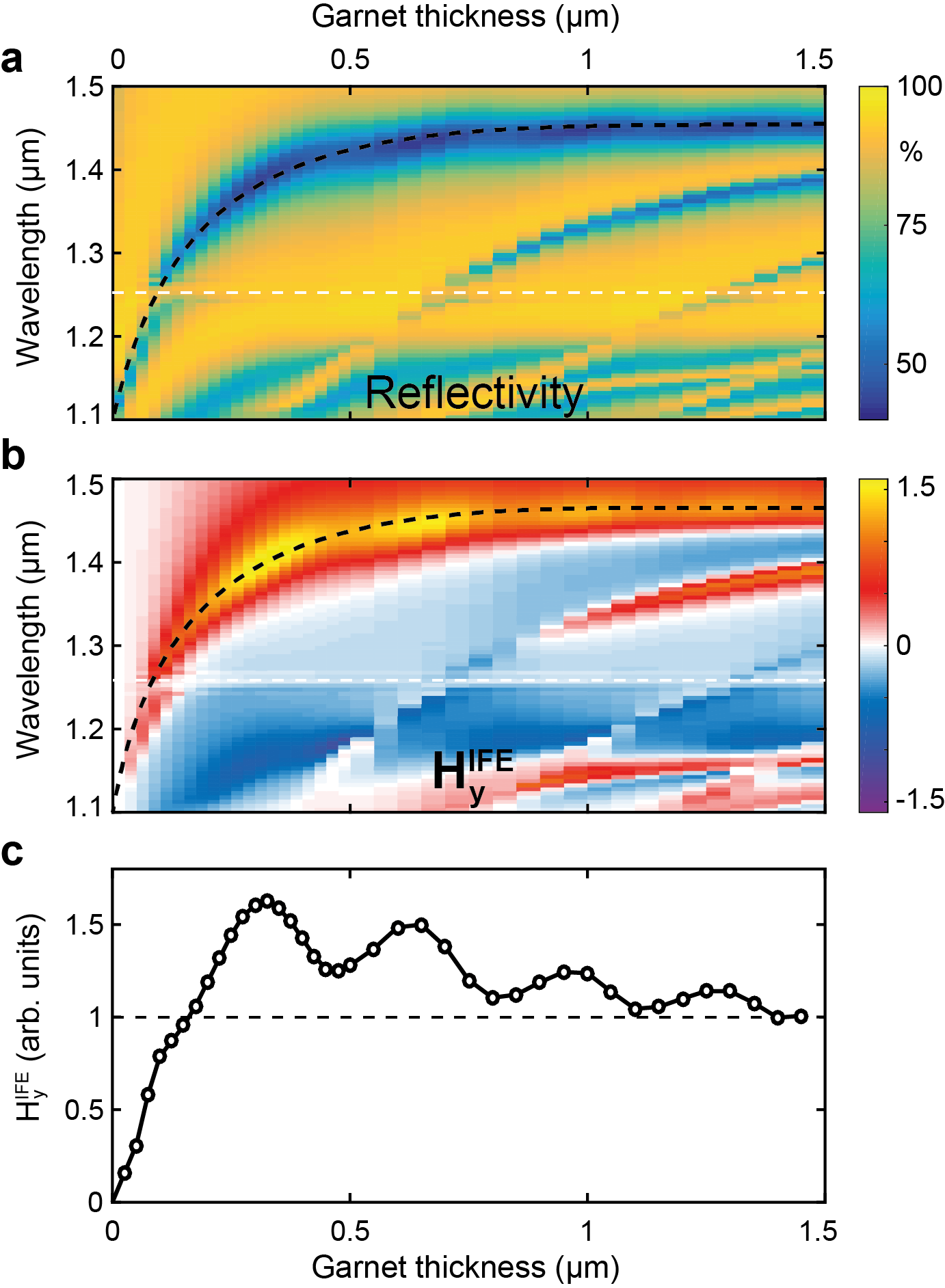}}
\caption{Calculated reflectivity (a) and integrated effective magnetic field (b) versus the wavelength and the garnet layer thickness. The SPP dispersion curves are shown with black (Au/garnet SPP) and white (Au/air SPP) dashed lines. (c) The integrated effective magnetic field of a Au/garnet SPP versus the garnet layer thickness.}
\label{SupFig3}
\end{figure*}

\end{document}